\documentclass{elsart}
\usepackage{graphicx}
\usepackage{amssymb}
\journal{Nucl. phys. A}
\begin{document}
\begin{frontmatter}

\title{Asymptotic normalization coefficient from the $^{12}$C($^7$Li,$^6$He)$^{13}$N reaction and the astrophysical $^{12}$C($p,\gamma$)$^{13}$N  reaction rate}
\author{Z. H. Li, }
\author{J. Su, }
\author{B. Guo, }
\author{Z. C. Li, }
\author{X. X. Bai, }
\author{J. C. Liu, }
\author{Y. J. Li, }
\author{S. Q. Yan, }
\author{B. X. Wang, }
\author{Y. B. Wang, }
\author{G. Lian, }
\author{S. Zeng, }
\author{E. T. Li, }
\author{X. Fang,}
\author{W. P. Liu, }
\author{Y. S. Chen, }
\author{N. C. Shu, }
\author{Q. W. Fan}
\address{China Institute of Atomic Energy, P. O. Box 275(46), Beijing 102413, PR China}

\begin{abstract}
Angular distribution of the $^{12}$C($^7$Li,$^6$He)$^{13}$N reaction
at $E$($^7$Li) = 44.0 MeV was measured at the HI-13 tandem
accelerator of Beijing, China. Asymptotic normalization coefficient
(ANC) of $^{13}$N $\rightarrow$ $^{12}$C + $p$ was derived to be
1.64 $\pm$ 0.11 fm$^{-1/2}$ through distorted wave Born
approximation (DWBA) analysis. The ANC was then used to deduce the
astrophysical $S(E)$ factors and reaction rates for direct capture
in $^{12}$C($p,\gamma$)$^{13}$N at energies of astrophysical
relevance.
\end{abstract}

\begin{keyword}
NUCLEAR REACTIONS $^{12}$C($^7$Li,$^6$He)$^{13}$N, $E$($^7$Li) =
44.0 MeV \sep measured $\sigma(\theta)$, DWBA analysis \sep deduced
asymptotic normalization coefficient \sep
$^{12}$C($p,\gamma$)$^{13}$N $E$ =
low \sep deduced astrophysical $S$-factor.\\
\PACS 25.70.Hi \sep 21.10.Jx \sep 25.40.Lw \sep 26.20.Cd
\end{keyword}
\end{frontmatter}

\section{Introduction}
In stellar evolution, the $^{12}$C($p,\gamma$)$^{13}$N reaction
plays a key role for the following reasons. Firstly, it is the first
reaction in the Carbon-Nitrogen-Oxygen (CNO) cycle which dominates
the energy production in stars with masses heavier than 1.5 M$_\odot
$ \cite{Rolf88,Math84}. Secondly, the $^{12}$C($p,
\gamma$)$^{13}$N($\beta^+$)$^{13}$C reactions can enhance the
$^{13}$C abundance \cite{Davi76}, and thus influence the
$^{12}$C/$^{13}$C ratio which is thought to be an important measure
for stellar evolution and nucleosynthesis \cite{Scho00}. Thirdly,
the supply of $^{13}$C by the $^{12}$C($p,
\gamma$)$^{13}$N($\beta^+$)$^{13}$C reactions is also important for
the $^{13}$C($\alpha, n$)$^{16}$O neutron source in the asymptotic
giant branch (AGB) stars. A recent calculation with parametric
one-zone nucleosynthesis showed that the more $^{13}$C supply is
needed at the end of the CNO cycle in solar-metallicity stars
\cite{Mow98}. In view of the above mentioned significance, it is
highly desired to carefully investigate the
$^{12}$C($p,\gamma$)$^{13}$N cross section at energies below 1.0 MeV
for the astrophysical interest.

The $^{12}$C($p,\gamma$)$^{13}$N reaction has been studied over a
wide energy range down to about $E_\mathrm{c.m.}$ = 70 keV
\cite{Bail50,Hall50,Lamb57,Rolf74,Bark80,Hebb60,Burt08,You63} since
1950. However, there exists obvious discrepancy in the experimental
data at the lower energies. In the even lower energy range, no
experimental data is available so far, and the astrophysical $S(E)$
factors can only be derived by the extrapolation which results in
uncertainty inevitably. In the energy range of $E_\mathrm{c.m.}$
$\leq$ 200 keV, the $^{12}$C($p,\gamma$)$^{13}$N reaction is
dominated by the tail of the s-wave capture into the broad $1/2^+$
resonance at $E_\mathrm{r}$ = 421 keV. Although the contribution
from direct capture is believed to be much smaller than that from
the resonance tail, the interference between the two processes can
lead to a considerable variation of $S(E)$ factors, since both of
them proceed via s-wave and then decay by $E1$ transitions. As a
result, the $S(E)$ factors either increase by the constructive
interference or decrease by the destructive one. Thus the reliable
experimental data on the direct capture are needed to derive the
$S(E)$ factors, particularly at energies of $E_\mathrm{c.m.}$ $\leq$
200 keV. A practicable scheme to deduce the $S(E)$ factors for the
direct capture in $^{12}$C($p,\gamma$)$^{13}$N is combining the
asymptotic normalization coefficient (ANC) of $^{13}$N $\rightarrow$
$^{12}$C + $p$ and R-matrix approach \cite{Bark91,Zhli06}, the ANC
can be deduced from the angular distribution of one proton transfer
reactions. The ($^7$Li,$^6$He) reactions are considered to be a
valuable spectroscopic tool because the shapes of their angular
distributions can be well reproduced by the distorted wave Born
approximation (DWBA) \cite{Kemp74}. Thus, the $^{12}$C($^7$Li,
$^6$He)$^{13}$N reaction is used to extract the nuclear ANC of
$^{13}$N $\rightarrow$ $^{12}$C + $p$ in the present work.

The $^{12}$C($^7$Li, $^6$He)$^{13}$N angular distribution was
measured at $E$($^7$Li) = 44.0 MeV. The spectroscopic factor and ANC
were derived based on DWBA analysis, and then used to calculate the
astrophysical $S(E)$ factors and rates of $^{12}$C($p,
\gamma$)$^{13}$N direct capture reaction at energies of
astrophysical interest with the R-matrix approach. We have also
computed the contribution from the resonant capture and the
interference effect between resonant and direct captures.

\section{Measurement of the $^{12}$C($^7$Li, $^6$He)$^{13}$N angular distribution}

The experiment was performed at the HI-13 tandem accelerator of
Beijing, China. A carbon target in thickness of 39.0 $\mu$g/cm$^2$
was bombarded with the 44.0 MeV $^7$Li beam in intensity of about
100 pnA. The $^6$He ions from the $^{12}$C($^7$Li, $^6$He)$^{13}$N
reaction were analyzed with the Q3D magnetic spectrograph. In order
to attain a good angular resolution, the solid angle for the
reaction products was set to be 0.23 msr. A two dimensional position
sensitive silicon detector (PSSD) was placed at the focal plane of
the spectrograph, which assured the full detection of $^6$He ions
emitted within the solid angle. During the measurement, a Faraday
cup placed at 0$^\circ$ in the reaction chamber was utilized to
record the beam current which served as the normalization standard
in determining the absolute cross section, while two independent
silicon detectors (SSDs) at $\pm$ 30$^{\circ}$ were used for both
monitoring the beam balance  and relative normalizing the measured
cross sections. The normalization was checked for several angles
from 23$^\circ$ to 40$^\circ$ with the elastic scattering data of
36.0 MeV $^7$Li on $^{12}$C \cite{Cobe76}.

\begin{figure}
\centering
\includegraphics[height=7 cm]{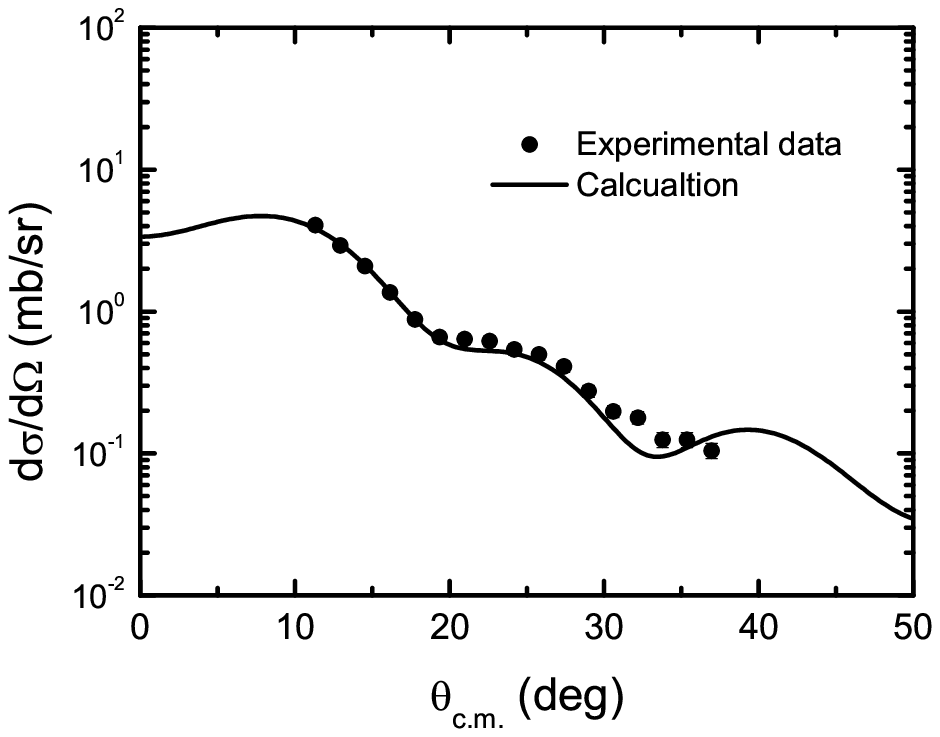}
\caption{\label{fig1}Angular distribution of the $^{12}$C($^7$Li,
$^6$He)$^{13}$N reaction at E($^7$Li) = 44.0 MeV.}
\end{figure}

The $^{12}$C($^7$Li, $^6$He)$^{13}$N differential cross sections
were measured in the angular range of 7$^{\circ}$ $\leq$
$\theta_\mathrm{lab}$ $\leq$ 23$^{\circ}$, corresponding to
11$^{\circ}$ $\leq \theta_\mathrm{c.m.} \leq$ 37$^{\circ}$. The
measured angular distribution is shown in Fig.~\ref{fig1}. The
experimental errors are from the uncertainties of statistics, target
thickness (5\%) and solid angle (6\%).

\section{Determination of the $^{13}$N Nuclear ANC}

The spins and parities of $^{12}$C and $^{13}$N ground states are
$0^{+}$ and $1/2^{-}$, respectively. The $^{12}$C($^7$Li,
$^6$He)$^{13}$N cross section is dominated by the (0$^+$,0)
$\rightarrow$ (1/2$^-$,1/2) transition, thus only $1p_{1/2}$ orbit
in $^{13}$N can be populated. If the reaction is peripheral, the
differential cross section can be expressed as
\begin{equation}\label{eq1}
({d\sigma \over d\Omega})_{exp} = (\frac{C_{^7Li}}{b_{^7Li}})^2
(\frac{C^{^{13}N}_{1,1/2}}{b^{^{13}N}_{1,1/2}})^2 ({d\sigma \over
d\Omega})_{DWBA},
\end{equation}
where $({d\sigma \over d\Omega})_{exp}$ and  $({d\sigma \over
d\Omega})_{DWBA}$ denote the measured and calculated differential
cross sections respectively. $C^{^{13}N}_{1,1/2}$ and $C_{^7Li}$ are
the nuclear ANCs of $^{13}$N $\rightarrow$ $^{12}$C + $p$ and $^7$Li
$\rightarrow$ $^6$He + $p$, $b^{^{13}N}_{1,1/2}$ and $b_{^7Li}$
being the single particle ANCs of the bound state protons in
$^{13}$N and $^7$Li, which can be calculated with the bound state
single particle wave function and Whittaker function at larger
radius. The ratio of C/b is so called spectroscopic factor. The
proton spectroscopic factor of $^7$Li was derived to be 0.41 $\pm$
0.05 in our previous work \cite{Zhli09}. The nuclear ANC of $^{13}$N
ground state can then be extracted by normalizing the theoretical
differential cross sections to the experimental data via Eq.
(\ref{eq1}).

The angular distribution of the $^{12}$C($^7$Li, $^6$He)$^{13}$N
reaction was calculated with the DWBA code FRESCO \cite{Thom88}. The
bound state wave function was obtained by solving the
Schr\"{o}dinger equation using a Woods-Saxon potential with standard
geometrical parameters ($r_0$ = 1.25 fm and $a$ = 0.65 fm), the
potential depth was adjusted so as to reproduce the observed binding
energy of the valence proton. The optical potential parameters for
both the entrance and exit channels were extracted by fitting the
angular distributions of elastic scattering with the code SFRESCO
\cite{Thom88}. The $^7$Li + $^{12}$C elastic scattering was measured
in the present work. Since there is no experimental data for $^6$He
elastic scattering on $^{13}$N , the angular distribution of $^6$Li
+ $^{13}$C elastic scattering at $E(^6\mathrm{Li})$ = 28.0 MeV
\cite{Bass72} was used to extract the potential parameters of the
exit channel. The extracted potentials are listed in
Table~\ref{tab1}.

\begin{table}
\caption{\label{tab1} Optical potential parameters used in the DWBA
calculations, where $U$, $W$ are in MeV, $r$ and $a$ in fm.}
\begin{center}
\begin{tabular}{cccccccc}
\hline \hline
Channel & $U_V$ & $r_{R}$ & $a_R$ & $W_V$ & $r_{I}$ & $a_{I}$ & $r_{c}$  \\
\hline
 $^7$Li + $^{12}$C & 194.55 & 0.50 & 0.82 & 7.66 & 1.31 & 0.75 & 1.30 \\
 $^6$He + $^{13}$N & 170.50 & 0.79 & 0.67 & 10.70 & 1.31 & 0.75 & 1.30 \\
\hline\hline
\end{tabular}
\end{center}
\end{table}

Generally, the spectroscopic factor or nuclear ANC is determined by
fitting the theoretical calculations to the experimental data at the
first peak in the angular distribution for the forward angles
\cite{Liu04}, since the experimental differential cross sections for
the backward angles are very sensitive to the inelastic coupling
effects and other high-order ones, which can not be well described
theoretically. In the DWBA calculation, the differential cross
sections at three forward angles were used to extract the ANC.
 The normalized angular distribution is also presented in
Fig.~\ref{fig1}. The nuclear ANC and the spectroscopic factor for
$^{13}$N $\rightarrow$ $^{12}$C + $p$ are deduced to be 1.64 $\pm$
0.11 fm$^{-1/2}$ and 0.64 $\pm$ 0.09, respectively. The
spectroscopic factors extracted by different experiments and the
theoretical values are listed in Tab.~\ref{tab2}. The spectroscopic
factor obtained in our work agrees with the theoretical ones
reported in Refs. \cite{Cohe67,Varm69} and the experimental results
given in Refs. \cite{Pear72,Koya76,Karb76,Nair75}. The Nuclear ANC
of $^{13}$N $\rightarrow$ $^{12}$C + $p$ from this work is in good
agreement with the value of 1.65 $\pm$ 0.20 fm$^{-1/2}$ extracted
from the $^{12}$C($^{10}$B,$^{9}$Be)$^{13}$N reaction \cite{Fern00}.

\begin{table}
\caption{\label{tab2} The theoretical and experimental proton
spectroscopic factors in the $^{13}$N ground state.}
\begin{center}
\begin{tabular}{cccc}
\hline \hline
$S_{^{13}N}$ & Experiments or theory  & Year & Reference \\
\hline
 0.61 & theory & 1967 & \cite{Cohe67} \\
 0.56 & theory & 1969 & \cite{Varm69} \\
 0.78 - 1.35 & $^{12}$C($d, n$) & 1970 &  \cite{Gang70} \\
 0.74  & $^{12}$C($d, n$) & 1971 &  \cite{Mutc71} \\
 0.53 $\pm$ 0.12 & $^{12}$C($d, n$) & 1972 &  \cite{Pear72} \\
 0.70 - 1.48 & $^{12}$C($^3$He, $d$) & 1969 &  \cite{Fort69} \\
 0.56 - 0.78 & $^{12}$C($^3$He, $d$) & 1976 &  \cite{Koya76} \\
 0.68 $\pm$ 0.12 & $^{12}$C($^3$He, $d$) & 1976 &  \cite{Karb76} \\
 0.81 $\pm$ 0.12 & $^{12}$C($^3$He, $d$) & 1979 &  \cite{Serc79} \\
 0.48 $\pm$ 0.12 & $^{12}$C($^3$He, $d$) & 1980 &  \cite{Pete80} \\
 1.34 & $^{12}$C($a, t$) & 1969 & \cite{Gail69} \\
 0.91 & $^{12}$C($a, t$) & 1972 & \cite{Haus72} \\
 0.72  & $^{12}$C($^7$Li, $^6$He) & 1979 &  \cite{Zell79} \\
 0.38 $\pm$ 0.05  & $^{12}$C($^7$Li, $^6$He) & 1986 &  \cite{Cook86} \\
 0.64 $\pm$ 0.09  & $^{12}$C($^7$Li, $^6$He) & 2008 &  present work \\
 0.25, 0.40 & $^{12}$C($^{10}$B, $^9$Be) & 1974 & \cite{Nair74} \\
 0.62 & $^{12}$C($^{14}$N, $^{13}$C) & 1975 & \cite{Nair75} \\
 0.29, 0.40 & $^{12}$C($^{16}$O, $^{15}$N) & 1979 & \cite{Prou79} \\
\hline\hline
\end{tabular}
\end{center}
\end{table}

\section{Astrophysical $S(E)$ factors of the $^{12}$C($p,\gamma$)$^{13}$N reaction}

Following the approach used in our previous work \cite{Zhli06}, the
astrophysical $S(E)$ factors of the $^{12}$C($p,\gamma$)$^{13}$N
reaction were calculated by the R-matrix method. For the radiative
capture reaction $B + b \rightarrow A + \gamma$, the cross section
to the state of nucleus A with the spin $J_f$ can be written as
\cite{Bark91,Zhli06}
\begin{equation}
\label{eq6} \sigma_{J_f}=\sum_{J_i} \sigma_{J_iJ_f},
\end{equation}

\begin{equation}
\label{eq7}
\sigma_{J_iJ_f}=\frac{\pi}{k^2}\frac{2J_i+1}{(2J_b+1)(2J_B+1)}\sum_{Il_i}|U_{Il_iJ_fJ_i}|^2,
\end{equation}

where $J_i$ denotes the total angular momentum of the colliding
nuclei B and b in the initial state, $J_b$ and $J_B$ are the spins
of nuclei b and B, and $I$, $k$ and $l_i$ are their channel spin,
wave number and orbital angular momentum in the initial state,
respectively. $U_{Il_iJ_fJ_i}$ is the transition amplitude from the
initial continuum state ($J_i,I,l_i$) to the final bound state
($J_f,I$). In the single-level, single-channel approximation, the
resonant amplitude for the capture into the resonance with energy
$E_{R_n}$ and spin $J_i$, and subsequent decay into the bound state
with the spin $J_f$ can be expressed as
\begin{equation}
\label{eq8}
U^R_{Il_iJ_fJ_i}=-ie^{i(\omega_{l_i}-\phi_{l_i})}\frac{[\Gamma^{J_i}_{bIl_i}(E)\Gamma^{J_i}_{\gamma
J_f}(E)]^{1/2}}{E - E_{R_n} + i \Gamma_{J_i}/2}.
\end{equation}
Here it is assumed that the boundary parameter is equal to the shift
function at resonance energy, and $\phi_{l_i}$ is the hard-sphere
phase shift in the $l_i$th partial wave,
\begin{equation}
\phi_{l_i}=\arctan\Big{[}\frac{F_{l_i}(k,r_c)}{G_{l_i}(k,r_c)}\Big{]},
\end{equation}
where $F^2_{l_i}$ and $G^2_{l_i}$ are the regular and irregular
 Coulomb functions, $r_c$ is the channel radius. The Coulomb phase factor $\omega_{l_i}$ is given by
\begin{equation}
\label{eq9} \omega_{l_i}=\sum_{n=1}^{l_i}\arctan(\frac{\eta_i}{n}),
\end{equation}
where $\eta_i$ is the Sommerfeld parameter.
$\Gamma^{J_i}_{bIl_i}(E)$ is the observable partial width of the
resonance in the channel $B$ + $b$, $\Gamma^{J_i}_{\gamma J_f}(E)$
is the observable radiative width for the decay of the given
resonance into the bound state with the spin $J_f$, and
$\Gamma_{J_i} \approx \sum\limits_I \Gamma^{J_i}_{bIl_i}$ is the
observable total width of the resonance level. The energy dependence
of the partial widths is determined by
\begin{equation}
\label{eq10}
\Gamma^{J_i}_{bIl_i}(E)=\frac{P_{l_i}(E)}{P_{l_i}(E_{R_n})}\Gamma^{J_i}_{bIl_i}(E_{R_n})
\end{equation}
and
\begin{equation}
\label{eq11} \Gamma^{J_i}_{\gamma
J_f}(E)=(\frac{E+\varepsilon_f}{E_{R_n}+\varepsilon_f})^{2L+1}\Gamma^{J_i}_{\gamma
J_f}(E_{R_n}),
\end{equation}
where $\Gamma^{J_i}_{bIl_i}(E_{R_n})$ and $\Gamma^{J_i}_{\gamma
J_f}(E_{R_n})$ are the experimental partial and radiative widths,
$\varepsilon_f$ is the proton binding energy of the bound state in
nucleus $A$, and L is the multipolarity of the gamma transition. The
penetrability $P_{l_i}(E)$ is expressed as
\begin{equation}
\label{eq12}
P_{l_i}(E)=\frac{kr_c}{F^2_{l_i}(k,r_c)+G^2_{l_i}(k,r_c)}.
\end{equation}

 The nonresonant amplitude can be calculated by
 \begin{eqnarray}
\label{eq13}
U^{NR}_{Il_iJ_fJ_i}&=&-(2)^{3/2}i^{l_i+L-l_f+1}e^{i(\omega_{l_i}-\phi_{l_i})}\frac{\mu_{Bb}^{L+1/2}}{\hbar
k}\Big{[}\frac{Z_be}{m^L_b}+(-1)^L\frac{Z_Be}{m^L_B}\Big{]}(k_\gamma
r_c)^{L+1/2}
\nonumber\\
&&  \times \sqrt{\frac{(L+1)(2L+1)}{L}}\frac{1}{(2L+1)!!}C_{J_fIl_f}F_{l_i}(k,r_c)G_{l_i}(k,r_c)\nonumber\\
&&\times W_{l_f}(2\kappa r_c)\sqrt{P_{l_i}}(l_i0L0|l_f0)
U(Ll_fJ_iI;l_iJ_f)J^\prime_L(l_il_f),
\end{eqnarray}
where $k_\gamma$ = $(E+\varepsilon_f)$/$\hbar c$ is the wave number
of the emitted photon. $C_{J_fIl_f}$ is the nuclear ANC of $^{13}$N
$\rightarrow$ $^{12}$C + $p$, and $l_f$ are the wave number and
relative orbital angular momentum of the bound state. $W_l(2\kappa
r)$ is the Whittaker hypergeometric function with $\kappa$ =
$\sqrt{2\mu_{Bb}\varepsilon_f}$. $(l_i0L0|l_f0)$ and
$U(Ll_fJ_iI;l_iJ_f)$ are the Clebsch-Gordan and Racha coefficients,
respectively. $J^\prime_L(l_il_f)$ is the integral expression
defined as
\begin{eqnarray}
\label{eq14} J^\prime_L(l_il_f)&=&
\frac{1}{r_c^{L+1}}\int^\infty_{r_c} dr \ r^L\frac{W_{l_f}(2\kappa
r)}{W_{l_f}(2\kappa
r_c)}\Big{[}\frac{F_{l_i}(k,r)}{F_{l_i}(k,r_c)}-\frac{G_{l_i}(k,r)}{G_{l_i}(k,r_c)}\Big{]}.
\end{eqnarray}

\begin{figure}
\centering
\includegraphics[height=7 cm]{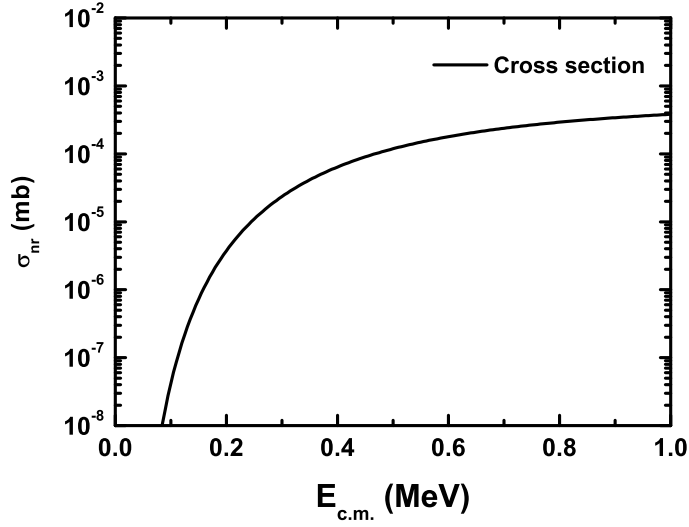}
\caption{\label{fig2}The non-resonant cross sections of the
$^{12}$C($p,\gamma$)$^{13}$N reaction computed with the ANC derived
from the present experiment.}
\end{figure}

The non-resonant amplitude contains the radial integral ranging only
from the channel radius $r_c$ to infinity since the internal
contribution is contained within the resonant part. Furthermore, the
R-matrix boundary condition at the channel radius $r_c$ implies that
the scattering of particles in the initial state is given by the
hard sphere phase. Hence, the problems related to the interior
contribution and the choice of incident channel optical parameters
do not occur. Therefore, the direct capture cross section only
depends on the ANC and the channel radius $r_c$. Using the
experimental ANC (1.64 $\pm$ 0.11 fm$^{-1/2}$) from the present
work, the non-resonant $^{12}$C($p,\gamma$)$^{13}$N cross sections
vs. $E_\mathrm{c.m.}$ were calculated, as shown in Fig~\ref{fig2}.
In the calculation, $r_c$ was taken to be 5.0 fm following the
previous works in Ref.~\cite{Zhli06} and \cite{Tang04}.

The astrophysical S-factor is related to the cross section by
\begin{equation}
\label{eq15}%
S(E)=E\sigma(E)\exp(E_{G}/E)^{1/2},
\end{equation}
where the Gamow energy $E_{G}=0.978Z^{2}_{1}Z^{2}_{2}\mu$ MeV, $\mu$
is the reduced mass of the system. Using the resonance parameters
($E_{R}$=421 keV, $\Gamma_{tot}(E_{R})=36.5 $ keV, and
$\Gamma_{\gamma}(E_{R})=0.67 $ eV) from Ref. \cite{Ajze70}, the
S-factors for direct and resonant captures can be then derived.

\begin{figure}
\centering
\includegraphics[height =8.0 cm]{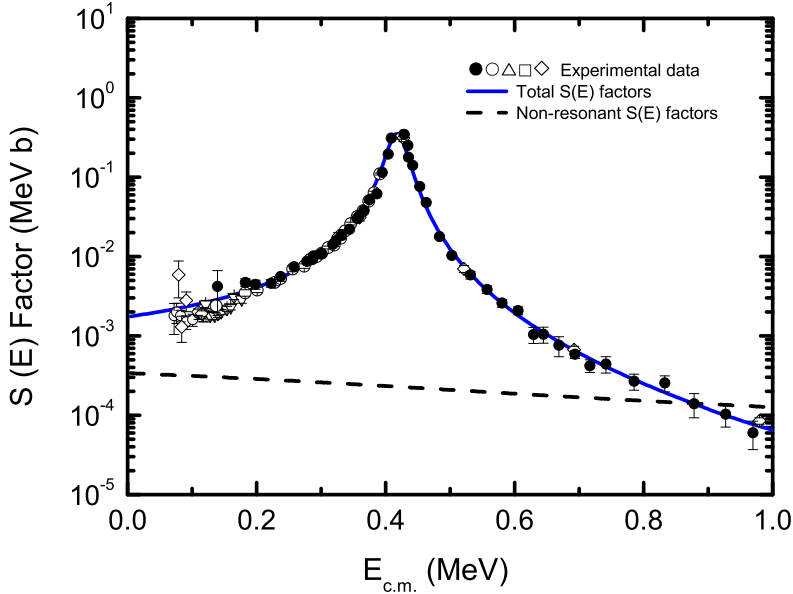}
\caption{\label{fig3}Astrophysical $S(E)$ factors as a function of
$E_{\mathrm{c.m.}}$ for the $^{12}$C($p,\gamma$)$^{13}$N reaction.
The dashed line and the solid line are the contributions from the
direct proton capture and the total $S(E)$ factors, respectively.
The experimental data are taken from Refs.
\cite{Bail50,Hall50,Lamb57,Rolf74,Burt08}}
\end{figure}

Since the incoming angular momentum ($s$-wave) and the multipolarity
($E1$) for the direct and resonant capture $\gamma$-radiation are
identical, there is an interference between the two capture
processes. Following the Ref.~\cite{Rolf74}, the total S-factor is
calculated by

\begin{equation}
\label{eq16} S_{tot}(E)=S_{dc}(E)+S_{res}(E) \pm
2[S_{dc}(E)S_{res}(E)]^{1/2}\cos(\delta_r),
\end{equation}
where $\delta_r$ is the resonance phase shift, given by

\begin{equation}
\label{eq17} \delta_r=\arctan\Big{[}{\Gamma_{p}(E) \over
2(E-E_{r})}\Big{]}.
\end{equation}

The experimental results from the direct measurement of
$^{12}$C($p,\gamma$)$^{13}$N reaction \cite{Rolf74} show that the
interference between the resonant and direct captures is
constructive below the resonance energy, and destructive above it.
Based on this interference pattern, the present total S-factors are
then obtained, as shown in Fig.~\ref{fig3}. One can see that the
present S-factors of $^{12}$C($p,\gamma$)$^{13}$N are in good
agreement with the experimental data from Ref.~\cite{Angu99} and the
references therein. Very recently, N. Burtebaev et al. \cite{Burt08}
measured the cross sections of the $^{12}$C($p, \gamma$)$^{13}$N
reaction at beam energies $E_p$ = 354, 390, 460, 463, 565, 750, and
1061 keV. They obtained ANC($^{13}$N) = 1.72 fm$^{-1/2}$,
$\Gamma_1^\gamma$ = 0.65 $\pm$ 0.07 eV and $\Gamma_1^p$ = 35.0 $\pm$
1.0 keV by fitting their experimental data with R-matrix approach.
Their results are in good agreement with ours.

\section{Astrophysical $^{12}$C($p,\gamma$)$^{13}$N reaction rates}

\begin{figure}[t]
\centering
\includegraphics[height =8.0 cm]{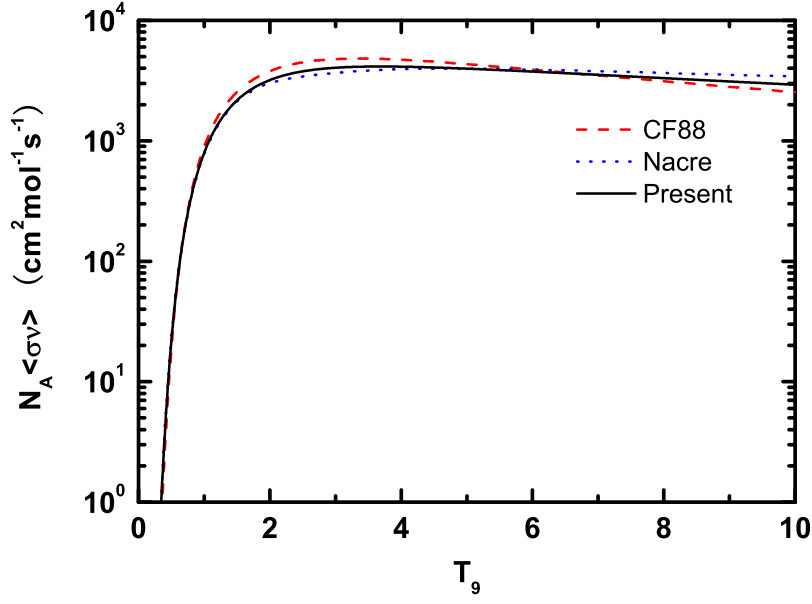}
\caption{\label{fig4} (Color online) The
$^{12}$C($p,\gamma$)$^{13}$N reaction rates from present work as
well as those from NACRE and CF88 compilations.}
\end{figure}

The astrophysical $^{12}$C($p,\gamma$)$^{13}$N reaction rate is
calculated with
\begin{eqnarray}
\label{eq18}%
N_A \langle\sigma v\rangle = N_A\big({8 \over \pi\mu}\big)^{1/2}{1
\over (k_BT)^{3/2}}\int^{\infty}_0 S(E)\exp\big[-({E_{G} \over
E})^{1/2}-{E \over k_BT}\big]dE,
\end{eqnarray}
where $v$ = $\sqrt{2E/\mu}$. $N_{A}$ and $k_B$ are Avogadro and
Boltzmann constants respectively. The updated reaction rates are
shown in Fig.~\ref{fig4}, together with the previous ones from NACRE
and CF88 compilations. The present reaction rates are larger than
that of NACRE and CF88 by about 20\% in the low temperature range of
$T_9$ $<$ 0.5, and very close to their average value in the
temperature range of 0.5 $\leq$ $T_9$ $\leq$ 10.0.

The total reaction rates as a function of temperature obtained in
our work are parameterized with an expression used in the
astrophysical reaction rate library REACLIB \cite{Thie87},

\begin{eqnarray}
\label{eq19}%
N_A \langle\sigma v\rangle&=& \exp(a_0 + a_1 T_9^{-1} + a_2
T_9^{-1/3} + a_3 T_9^{1/3} + a_4 T_9 \nonumber\\&+& a_5 T_9^{5/3} +
a_6 \ln T_9)+\exp(b_0 + b_1 T_9^{-1}+b_2 T_9^{-1/3} \nonumber\\&+&
b_3 T_9^{1/3}+ b_{4} T_9 + b_{5} T_9^{5/3} + b_{6} \ln T_9).
\end{eqnarray}

\begin{table}
\caption{\label{tab3} The fitting parameters of the
$^{12}$C($p,\gamma$)$^{13}$N reaction rates.}
\begin{center}
\begin{tabular}{cccc}
\hline \hline parameters & value & parameters & value \\
\hline
$a_0$ &  1.1397E+01 & $b_{0}$ &  4.0200E+01 \\
$a_1$ & -2.9913E-02 & $b_{1}$ &  2.1886E+00 \\
$a_2$ & -1.1507E+01 & $b_{2}$ & -1.5188E+02 \\
$a_3$ &  7.6988E+00 & $b_{3}$ &  1.1973E+02 \\
$a_4$ & -3.4841E+00 & $b_{4}$ & -3.7889E+00 \\
$a_5$ &  3.8122E-01 & $b_{5}$ &  1.2086E-01 \\
$a_6$ &  4.5641E-02 & $b_{6}$ & -8.1517E+01 \\
\hline\hline
\end{tabular}
\end{center}
\end{table}

The value of fit parameters $a_{0 - 6}$ and $b_{0 - 6}$ are listed
in Tab.~\ref{tab3}, and the fitting errors are less than 1\%
in the temperature range of 0.01 $\leq T_{9} \leq $ 10.\\

\section{Conclusion and discussion}
The $^{12}$C($p,\gamma$)$^{13}$N reaction is of considerably
astrophysical interest. The cross sections in the energy range of
astrophysical relevance is very small and difficult to be measured
directly. In the energy range of $E_\mathrm{c.m.}$ $\leq$ 200 keV,
the $^{12}$C($p,\gamma$)$^{13}$N cross section depends on the tail
of the s-wave capture into the broad $1/2^+$ resonance at
$E_\mathrm{r}$ = 421 keV, the direct capture to the ground state and
their interference. The determination of the resonant parameters for
the $1/2^+$ state and the ANC of $^{13}$N ground state is helpful in
extrapolating the experimental data at high energies down to the
energies of astrophysical interest (around 25 keV).

 We measured the angular distribution of the $^{12}$C($^7$Li, $^6$He)$^{13}$N
reaction at $E_\mathrm{^7Li}$ = 44.0 MeV, and deduced the nuclear
ANC and spectroscopic factor for the $^{13}$N ground state. The
astrophysical S-factors and reaction rates of $^{12}$C($p,
\gamma$)$^{13}$N are then extracted with the R-matrix approach. The
S-factor at 25.0 keV is found to be 1.87 $\pm$ 0.13 keV$\cdot$b. The
result is consistent with that of 1.75 $\pm$ 0.22 keV$\cdot$b
obtained by Burteaev et al. \cite{Burt08}, and slightly higher than
the earlier values of 1.45 $\pm$ 0.20 keV$\cdot$b, 1.54 $\pm$ 0.08
keV$\cdot$b and 1.33 $\pm$ 0.15 keV$\cdot$b reported in Refs.
\cite{Rolf74,Bark80,Hebb60}. In order to clarify the deviation, the
further study of the resonant parameters of the $1/2^+$ state is
needed.

\begin{ack}
This work is supported by the National Basic Research Programme of
China under Grant No. 2007CB815003, the National Natural Science
Foundation of China under Grant Nos. 10675173, 10705053 and
10735100.
\end{ack}

\end{document}